\documentclass[prb]{revtex4}
\usepackage{graphicx}

\begin{document}

\begin{large}

\begin{center}
\begin{Large}
{\bf{ Surface phase transitions in one-dimensional channels
arranged in a triangular cross-sectional structure: Theory and
Monte Carlo simulations }}
\end{Large}
\end{center}

\vspace{0.8cm}

\begin{center}
P. M. Pasinetti$^{1}$, F. Rom\'a$^{1,2}$, J. L. Riccardo$^{1}$ and
A. J. Ramirez-Pastor$^{1,\dag}$
\end{center}

\begin{center} {
1. Departamento de F\'{\i}sica, Universidad Nacional de San Luis, \\
 CONICET, Chacabuco 917, 5700 San Luis, Argentina \\
 2. Centro At\'omico Bariloche, CONICET Av. Bustillo 9500, \\
8400 S. C. de Bariloche, Argentina \\}
\end{center}

\begin{center} Abstract \end{center}
Monte Carlo simulations and finite-size scaling analysis have been
carried out to study the critical behavior in a submonolayer
lattice-gas of interacting monomers adsorbed on one-dimensional
channels arranged in a triangular cross-sectional structure. The
model mimics a nanoporous environment, where each nanotube or unit
cell is represented by a one-dimensional array. Two kinds of
lateral interaction energies have been considered: $1)$ $w_L$,
interaction energy between nearest-neighbor particles adsorbed
along a single channel and $2)$ $w_T$, interaction energy between
particles adsorbed across nearest-neighbor channels. We focus on
the case of repulsive transversal interactions ($w_T > 0$), where
a rich variety of structural orderings are observed in the
adlayer, depending on the value of the parameters $k_BT/w_T$
(being $k_B$ the Boltzmann constant) and $w_L/w_T$. For
$w_L/w_T=0$, successive planes are uncorrelated, the system is
equivalent to the triangular lattice and the well-known $(\sqrt{3}
\times \sqrt{3})$ $\left[(\sqrt{3} \times \sqrt{3})^*\right]$
ordered phase is found at low temperatures and a coverage,
$\theta$, of $1/3$ $[2/3]$. In the more general case ($w_L/w_T
\neq 0$), a competition between interactions along a single
channel and a transverse coupling between sites in neighboring
channels allows to evolve to a three-dimensional adsorbed layer.
Consequently, the $(\sqrt{3} \times \sqrt{3})$ and $(\sqrt{3}
\times \sqrt{3})^*$ structures ``propagate" along the channels and
new ordered phases appear in the adlayer. Each ordered phase is
separated from the disordered state by a continuous order-disorder
phase transition occurring at a critical temperature, $T_c$, which
presents an interesting dependence with $w_L/w_T$. The Monte Carlo
technique was combined with the recently reported Free Energy Minimization Criterion Approach (FEMCA),
[F. Rom\'a et al., Phys. Rev. B, 68, 205407, (2003)], to predict
the critical temperatures of the order-disorder transformation.
The excellent qualitative agreement between simulated data and
FEMCA results
 allow us to interpret the physical meaning of the mechanisms underlying
the observed transitions.

\vspace{1mm} \noindent {\it {\bf Keywords:} Lattice-Gas Models,
Phase Transitions, Monte Carlo Simulation} \vspace{1mm}

\noindent $\dag$ To whom all correspondence should be addressed.

\newpage

\section{Introduction} \label{Introduction}

Lattice-gas models have been extensively investigated in the last
decades because they provide a theoretical framework for the
description of many physical, chemical and biological systems.
The adsorption thermodynamics and the understanding of surface phenomena
have been greatly benefited from the development of these models
\cite{Fowler,Hill,Stanley}.
In this sense, the more recognized examples are the Langmuir adsorption
model \cite{Hill,Langmuir} and the Ising model of magnetism \cite{Hill,Ising,Kramers,Montroll,Onsager}.
More recently, a number of contributions have been devoted to the study
of adsorption of gases on solid surface \cite{Binder,Landau,Phares,PRE2,Patry,JCP1PRE1,Nitta,Rudzi,Boro1,Boro2,PRB3,SS3LANGMUIR5,LANGMUIR6,PRB4}. These papers have included
the effects of lateral interactions \cite{Binder,Landau,Phares,PRE2,Patry},
surface heterogeneity \cite{JCP1PRE1},
multisite occupation \cite{Phares,Nitta,Rudzi,Boro1,Boro2,PRB3,SS3LANGMUIR5,LANGMUIR6,PRB4}, etc.
Among them, A. J. Phares et al. \cite{Phares}.
have studied the structural orderings occuring in a wide variety of experimental
and theoretical systems and its influence on the corresponding phase diagrams.

Recently, the advent of modern techniques for building single and
multiwalled carbon nanotubes
\cite{Ijima,Ijima1,Bethune,Ajayan,Ebbensen} has considerably
encouraged the investigation of the gas-solid interaction
(adsorption and transport of simple and polyatomic adsorbates) in
such a low-dimensional confining adsorption potentials. The design
of carbon tubules, as well as of synthetic zeolites and
aluminophosphates such as $AlPO_4-5$ (Ref. \cite{Martin}) having
narrow channels, literally provides a way to the experimental
realization of quasi one-dimensional adsorbents. Many studies on
conductivity, electronic structure, mechanical strength, etc. of
carbon nanotubes are being currently carried out. However the
amount of theoretical and experimental work done on the
interaction and thermodynamics of simple gases adsorbed in
nanotubes is still limited
\cite{Gubbins,Boutin,Lachet,Maris,Martin1}. These papers predict a
complex behavior. At low temperatures, Radhakrishnan et al.
\cite{Gubbins} have reported, by using molecular simulations, a
phase transition occurring between two different density states,
one corresponding to a low density (``gas'') and the other to a
higher density (``liquid'') phase. The authors propose that this
transition is induced by the interactions with the molecules that
are in neighboring channels. A different type of transition was
found by Boutin et al. \cite{Boutin} and Lachet et al.
\cite{Lachet} at intermediate temperatures (much higher
temperature than the phase transition considered by Radhakrishnan
et al.). The molecular simulations of Refs. \cite{Boutin,Lachet}
show stepped isotherms. The step was only observed using a model
that includes both two-body and three-body lateral interactions,
whereas a model with only two-body interactions did not show such
a step. Later, this transition has been confirmed by experimental
results by Martin et al. \cite{Martin1}. The step was attributed
to local rearrangement of the adsorbed phase. Other interesting
features of the carbon nanotubes have been recently studied.
Namely, Urban et al. \cite{Urban} studied the properties of an
$Ar$ film adsorbed on the external surface of a bundle of carbon
nanotubes; and Calbi and Riccardo \cite{Calbi} investigated the
presence of adsorption sites and energy barriers near the ends of
carbon nanotube bundles to determine their consequences on gas
adsorption in the interstitial channels between the tubes.

For theoretical purposes, adsorption in a narrowest nanotube can
be treated in the one-dimensional lattice-gas approach
\cite{PRB3,PRE2}. This is, of course, an approximation to the
state of real adsorbata in nanotubes, which is justified because
thermodynamics and transport coefficient can be analytically
resolved in these conditions. Many studies have been performed of
specific geometries and specific adsorbate-substrate combinations
\cite{Gelb}. In recent papers \cite{Cole1,Cole}, Trasca et al.
presented simple model of a nanoporous environment. That is a
lattice-gas model with two kinds of sites. One was a
one-dimensional line of sites, which the authors called ``axial''
sites, surrounded by a set of ``cylindrical shell'' sites. The
system was evaluated by using mean-field theoretical approach
\cite{Cole1} and Monte Carlo (MC) simulations \cite{Cole}. This
work represents an effort in that direction. Here, we study a
simplified lattice-gas model, which can help us to establish
criteria to characterize more complex experimental systems. In
this model, each nanotube has been represented by a
one-dimensional chain. These chains were arranged in a triangular
structure. We included longitudinal interactions between
nearest-neighbor particles adsorbed along a single channel, $w_L$,
and transversal energy between particles adsorbed across
nearest-neighbor channels, $w_T$. The phase behavior depends on
the values of these various energies, especially on the attractive
or repulsive character of the interaction. In previous work, low
temperature calculations of configurational entropy of the adlayer
allowed us to identify a wide variety of structural orderings
\cite{entropy}. Later, the influence of such structural orderings
on interesting properties as adsorption isotherm and heat of
adsorption was analyzed \cite{isotherm}. The present article goes
a step further, studying the critical behavior of the system via
MC simulation, finite-size scaling analysis and the recently
reported Free Energy Minimization Criterion Approach (FEMCA)
\cite{PRB4}. For this purpose, the critical temperature $T_c$
characterizing the transition from the disordered state to the
ordered phase is obtained as a function of the ratio $w_L/w_T$.

The paper is organized as follows: In Section II we describe the
lattice-gas model, the simulation scheme and we present the
behavior of $T_c(w_L/w_T)$, obtained by using MC method. In
Section III we present the theoretical approach (FEMCA) and
compare the MC results with the theoretical calculations. Finally,
the general conclusions are given in Section IV.

\section{Lattice-gas model and Monte Carlo simulation scheme}
\subsection{The model}

We consider the adsorption of monomers on a simple model of a
nanoporous environment. In this model, each nanotube or unit cell
has been represented by a one-dimensional (1D) line of $L$
adsorptive sites, with periodical boundary conditions. These
chains were arranged in a triangular structure of size $R \times
R$ and periodical boundary conditions. Under these condition all
lattice sites are equivalent hence border effects will not enter
our derivation. The energies involved in the adsorption process
are three:
\begin{itemize}
\item[1).] $\varepsilon_0$, interaction energy between a particle
and a lattice site. \item[2).] $w_{L}$, interaction energy between
adjacent occupied axial sites. \item[3).] $w_{T}$, interaction
energy between particles adsorbed on nearest-neighbor transverse
sites.
\end{itemize}

Thus, the resulting substrate was an anisotropic three-dimensional
array of $M=L \times R \times R$ adsorption sites, where each site
was surrounded by two ``axial" sites along the nanotube's axis and
six ``transverse" sites belonging to nearest-neighbor unit cells
(see Fig. 1 in Ref.\cite{entropy}).

In order to describe the system of $N$ molecules adsorbed on $M$
sites at a given temperature $T$, let us introduce the occupation
variable $c_{i,j,k}$ which can take the following values:
$c_{i,j,k}=0$ if the corresponding site $(i,j,k)$ is empty and
$c_{i,j,k} = 1$ if the site is  occupied by an adatom. Then, the
Hamiltonian of the system is given by,
\begin{equation}
H = w_L \sum_{{\langle i,j,k; i',j',k' \rangle}_L} c_{i,j,k}
c_{i',j',k'} + w_T \sum_{{\langle i,j,k; i',j',k' \rangle}_T}
c_{i,j,k} c_{i',j',k'} + (\varepsilon_0 - \mu) \sum_{i,j,k}^M
c_{i,j,k} \label{ham}
\end{equation}
where ${{\langle i,j,k; i',j',k' \rangle}_L}$ (${{\langle i,j,k; i',j',k' \rangle}_T}$) represents
pairs of $NN$ axial (transverse) sites and $\mu$ is the chemical potential.

\subsection{Monte Carlo simulations}\label{simulation}

The lattice-gas was generated fulfilling the following conditions:

\begin{itemize}
\item The sites were arranged in a structure of size $M=L \times R
\times R$, with conventional periodic boundary conditions. \item
Due to the surface was assumed to be homogeneous, the interaction
energy between the adparticles and the atoms of the substrate,
$\varepsilon_0$,  was neglected for sake of simplicity. \item
Repulsive transversal lateral interactions were considered, where
a rich variety of ordered phases are observed in the planes. \item
Repulsive and attractive longitudinal lateral interactions were
used. \item Appropriate values of $L$ and $R$ were used in such a
way that the adlayer structures at critical regime are not
perturbed.
\end{itemize}

In order to study the critical behavior of the system, we have
used an efficient exchange MC or simulated tempering method
\cite{Hukushima,Earl} and finite-size scaling analysis
\cite{Fisher,Privman,Binder2}. As in Ref. \cite{Hukushima}, we
build a compound system which consists of $m$ non-interacting
replicas of the system concerned. The $m$-th replica is associated
with the temperature $T_m$ [or $\beta_m=1/\left(k_B T_m\right)$,
being $k_B$ the Boltzmann constant]. In other words, each replica
is in contact with its own heat bath having different temperature.
Under these conditions, the algorithm to carry out the simulation
process is the following:

\begin{itemize}
\item[1)] The compound system of $m$ replicas is generated. For
this purpose, each replica is simulated simultaneously and
independently as canonical ensemble for $n_1$ MC steps by using a
standard importance sampling MC method
\cite{Binder2,Nicholson,Metropolis}. In order to determine the set
of temperatures, $\{T_m \}$ ($\{\beta_m \}$), we set the highest
temperature, $T_{max}$ ($\beta_{min}$), in the high temperature
phase where relaxation (correlation) time is expected to be very
short and there exists only one minimum in the free energy space.
On the other hand, the lowest temperature, $T_{min}$
($\beta_{max}$), is somewhere in the low temperature phase whose
properties we are interested in. Finally, the difference between
two consecutive temperatures is set as $\left(T_{max} - T_{min}
\right)/(m-1)$ (equally spaced temperatures).

\item[2)] Interchange vacancy-particle. The procedure is as follows:
\begin{itemize}
\item[2.1)] One of the $m$ replicas is randomly selected.
\item[2.2)] An occupied site and an empty site, both belonging to the
replica chosen in 2.1), are randomly selected and their positions are established.
\item[2.3)] By using a standard Kawasaki algorithm \cite{Kawasaki}, an attempt is made
to interchange the occupancy state of the sites chosen in step 2.2).
\end{itemize}

\item[3)] Exchange of two configurations $X_m$ and $X_{m'}$,
corresponding to the $m$-th and $m'$-th replicas is tried and
accepted with the probability $W\left(X_m,\beta_m|
X_{m'},\beta_{m'}\right)$. In general, the probability of
exchanging configurations of the $m$-th and $m'$-th replicas is
given by \cite{Hukushima},
\begin{equation}
W\left(X_m,\beta_m| X_{m'},\beta_{m'}\right)=\left\{
\begin{array}{cc}
1 & {\rm for}\ \ {\Delta<0} \\
\exp(-\Delta)  & {\rm for}\ \ {\Delta>0}
\end{array}
\right.
\end{equation}
where $\Delta=\left( \beta_m - \beta_{m'} \right)\left[ H(X_{m'})
- H(X_{m}) \right]$. As in Ref. \cite{Hukushima}, we restrict the
replica-exchange to the case $m \leftrightarrow m+1$.


\item[4)] Repeat from step 2) $m \times M $ times. This is the
elementary step in the simulation process or Monte Carlo step
(MCS).

\end{itemize}

The procedure 1)-4) is repeated for all lattice's sizes. For each
size, the equilibrium state can be well reproduced after
discarding the first $n_2$ MCS. Then, averages are taken over
$n_{MCS}$ successive MCS. The canonical expectation value of a
physical quantity $A$ is obtained in the usual way as follows:
\begin{equation}
{\langle A\rangle}_{\beta_m}=\frac{1}{n_{MCS}} \sum_{t=1}^{n_{MCS}} A \left[X_m(t)\right]
\end{equation}

All calculations were carried out using the parallel cluster BACO
of Universidad Nacional de San Luis, Argentina. This facility
consists of 60 PCs each with a 3.0 GHz Pentium-4 processor.


As it is standard for order-disorder phase transitions, a related
order parameter was defined. In particular, at $\theta=1/3$
[$2/3$] (being $\theta \equiv N/M$ the surface coverage), a
$(\sqrt{3} \times \sqrt{3})$ [$(\sqrt{3} \times \sqrt{3})^*$]
ordered structure is formed in the planes below the critical
temperature. Depend on the sign of the longitudinal interactions,
the order is propagated to all planes. For repulsive $w_L$,
adatoms avoiding configurations with nearest-neighbor interactions
 order along the channels in a structure of alternating particles separated
by empty sites. On the other hand, attractive monomer-monomer
longitudinal interactions favor the formation of pairs of
nearest-neighbor adsorbed particles along the nanotubes. The
resulting structures are shown in Figs. 5 and 8 of
Ref.\cite{entropy}.



For the case of repulsive longitudinal interactions, Fig. 1 a)
shows two successive planes, $k$ and $k+1$, for one possible
configuration of the phase appearing at critical regime and
$\theta=1/3$. Due to the periodic boundary conditions the
degeneracy of this ``local phase" is equal to six. These
configurations allow us to decompose the ``local lattice" into six
different sublattices [see Figs. 1 b)-c)] \cite{foot1}. The
coverage on each sublattice is denoted as $\theta_s (s=1,...,6)$.
In this way, an ``local order parameter", $\varphi_k$, can be
defined as

\begin{equation}
\varphi_k = \sum_{s,t; s \neq t} |\theta_s-\theta_t| \label{fidim}
\end{equation}

\noindent where we sum the differences (in absolute value) between the coverage corresponding
to two sublattices.

When the system is disordered $(T>T_c)$, all sublattices are
equivalents and the order parameter is minimum. However, when a
configuration of the local phase appears at low temperature
$(T<T_c)$, this is allocated on a sublattice. Let us suppose that
this configuration lies on the sublattice $s$. Then, the coverage
$\theta_s$ is maximum ($\theta_s=1$) and the coverage of the rest
of the sublattices is zero or minimum. Consequently, $\varphi_k$
is also maximum.

On the basis of $\varphi_k$, the generalized order parameter, $\varphi$, can be written as,

\begin{equation}
\varphi = A\sum_{k=0}^{L} \varphi_k \label{fi}
\end{equation}
where $A$ is a normalization factor. The definition (\ref{fi})
is computationally convenient and $\varphi$ appears as a good
order parameter evidencing the order-disorder phase transition.

In a similar way, it is possible to define the order parameter corresponding
to $\theta=2/3$ and repulsive longitudinal interactions.

For attractive longitudinal interactions, it is not appropriate to
define local sublattices. In this case, each sublattice $s
(s=1,...,3)$ lies on the total lattice (see Fig. 2) and $\varphi$
can be easily defined as $\varphi =
|\theta_1-\theta_2|+|\theta_1-\theta_3|+|\theta_2-\theta_3|$

Now, the reduced fourth-order cumulant, $U_{R[L]}$, introduced by
Binder \cite{Binder2} and related to the order parameter, can be
calculated as:

\begin{equation}
U_{R[L]}(T) = 1 -\frac{\langle \varphi^4\rangle_T} {3\langle
\varphi^2\rangle_T^2} \label{cum}
\end{equation}

\noindent where $U_R$ [$U_L$] represents the cumulant obtained by
variable $R$ [$L$] and fixed $L$ [$R$]. The thermal average
$\langle ... \rangle_T$, in all the quantities, means the time
average throughout the MC simulation.

The standard theory of finite-size scaling
\cite{Fisher,Privman,Binder2} allows for various efficient routes
to estimate $T_c$ from MC data. One of these methods, which will
be used here, is from the temperature dependence of $U_{R[L]}(T)$,
which is independent of the system size for $T=T_c$. In other
words, $T_c$ is found from the intersection of the curve
$U_{R[L]}(T)$ for different values of $R[L]$, since
$U_{R[L]}(T_c)=$ constant.

\subsection{Computational results}\label{comres}

The thermodynamic properties of the present model have been
investigated by means of the computational scheme described in the
previous section. As a consequence of the equivalence
particle-vacancy, the critical behavior at $\theta = 2/3$ is as at
$\theta=1/3$. Then, we restrict our calculations to $\theta=1/3$.
In addition, we set $w_T=1$ and vary $w_L/w_T$ from $-1$ to $1$
($-1 \leq w_L/w_T \leq 1$).

In order to understand the basic phenomenology, we consider in the
first place null longitudinal interactions ($w_L/w_T=0$). In this
particular case, successive planes are uncorrelated and the system
is equivalent to the well-known triangular lattice. The value
obtained of $k_BT_c/w_T=0.3354(1)$ confirms this arguments and
validates the MC scheme \cite{Metcalf,Kinzel,Chin,Triang}. The
data are not shown here for brevity.

Hereafter, we discuss the behavior of the critical temperature as
a function of $w_L/w_T$. We start with the case of attractive
longitudinal interactions. As an example, Fig. 3 illustrates the
reduced four-order cumulants plotted versus $k_BT/w_T$ for
$w_L/w_T = -1$. From their intersections one gets the estimation
of the critical temperature. The lattice sizes used in the
simulation \cite{foot2} are compiled in Table I along with the
values of the parameters in the simulated tempering runs. In the
figure, the critical temperature is obtained from the curves of
$U_R(T)$ (calculated for different values of $R$ and fixed $L$).
The resulting value, $k_BT_c/w_T=0.7817(1)$, agrees very well with
previous determinations reported in the literature \cite{entropy}.
In Ref. \cite{entropy}, a value $k_BT_c/w_T \approx 0.76$ was
obtained from the inflection on the function $s(T)$, being $s(T)$
the configurational entropy per site of the adlayer as a function
of the temperature. Due to the finite-size scaling technique and
the simulation procedure used in this contribution, the estimation
of $T_c$ in the present work is expected to be more accurate than
that reported previously.

The study was extended to repulsive longitudinal interactions.
Fig. 4 shows the data for a typical case ($w_L/w_T=1$), resulting
$k_BT_c/w_T=0.6098(5)$, in well agreement with the value obtained
in Ref. \cite{entropy} ($k_BT_c/w_T \approx 0.59$). As indicated
in Fig. 3, the parameters of the simulation are listed in Table I.

Due to the presence of anisotropy (the couplings are taken to be
different in the different lattice directions), it is expected
that the correlation functions in transversal and longitudinal
directions may be governed by correlation lengths diverging with
different critical exponents \cite{Barouch,BinderWang}. However,
it is worth pointing out that we do not assume any particular
value of the critical exponents
 for the transitions analyzed here in order to
calculate their critical temperatures, since the analysis rely on
the order parameter cumulant's properties \cite{foot3}. In
addition, the procedure shown in Figs. 4 and 5 was repeated for
the curves of $U_L(T)$, which were obtained for variable $L$ and
fixed $R$. As an example, Fig. 6 presents the results obtained for
the case $w_L/w_T = -1$. As it is expected, identical results
(within numerical errors) are obtained in both ways.

Finally, the calculations were carried out for $w_L/w_T=-0.75,
-0.50, -0.25, 0.00, 0.25, 0.50$ and $0.75$ and the results were
collected in Table II \cite{foot4}. As it can be observed, the
critical temperature presents an non-trivial behavior as a
function of $w_L/w_T$. An understanding of the dependence of
$k_BT_c/w_T$ on $w_L/w_T$ can be developed by following the subtle
interdependence of energetic and entropic cost necessary to alter
the ordered phase. This will be discussed in Section III.

\section{Theoretical approach: Free Energy Minimization Criterion}

Hereafter, we will use FEMCA \cite{PRB4} in order to discuss
the dependence of $k_BT_c/w_T$ vs. $w_L/w_T$ obtained
from MC simulation.

In a closed system of adsorbed particles with repulsive
interactions, the phase transition occurring in the adsorbate is a
continuous (second-order) phase transition. In other words, the
entropy, $S$, varies continuously from a completely ordered state
(when $T \rightarrow 0$) to a disordered state (when $T
\rightarrow \infty$). Around $T_c$, $S$ changes abruptly (but
continuously) \cite{LANGMUIR6}. Then, it's possible to analyze the
phase transition taking into account the Helmholtz free energy,
$F=E-TS$ (being $E$ the mean energy), in the two extreme states
(maximum order and maximum disorder). Accordingly,

\begin{equation}
F_{\infty} = \lim_{T \rightarrow \infty} F \  \  \  \ {\rm and}
\  \  \  \  F_{0} = \lim_{T \rightarrow 0} F  \label{fefinf}
\end{equation}

then

\begin{equation}
F_{\infty} << F_{0} \  \ \Rightarrow \  \ T > T_c  \label{fet1}
\end{equation}

\begin{equation}
F_{\infty} >> F_{0} \  \ \Rightarrow \  \ T < T_c  \label{fet2}
\end{equation}

\begin{equation}
F_{\infty} = F_{0} \  \ \Rightarrow \  \ T \approx T_c  \label{fet3}
\end{equation}

The last equation allows to determine $T_c$. This calculation is
not exact due to the system does not pass from an extreme order to
an extreme disorder. There exist intermediate states between the
two extreme states. However, as we will show in the following
analysis, the eq. (\ref{fet3}) provides a very good approximation
for $T_c$. Interested readers are referred to Ref. \cite{PRB4} for
a more complete description of FEMCA.

In general, for a system of interacting particles at temperature $T$ results:

\begin{equation}
f_0 = e_0 -T s_0    \  \  \  \ {\rm and} \  \  \  \    f_{\infty}
= e_{\infty} -T s_{\infty}    \label{fef0}
\end{equation}
where $e$ and $s$ represent the mean energy per site and the
entropy per site in the thermodynamical limit, respectively,

\begin{equation}
e = \lim_{M \rightarrow \infty} \frac{E}{M} \  \  \  \ {\rm and} \
\  \  \  s = \lim_{M \rightarrow \infty} \frac{S}{M}  \label{fe1}
\end{equation}

If $f_0=f_{\infty}$, this is
\begin{equation}
e_0 -T s_0  =  e_{\infty} -T s_{\infty}
\end{equation}
then $T \approx T_c$ and
\begin{equation}
T_c  \approx \frac{\Delta e}{\Delta s} =
\frac{e_{\infty}-e_0}{s_{\infty}-s_0} \label{fetc1}
\end{equation}
From eq. (\ref{fetc1}), it is possible to calculate the critical
temperature and to interpret the dependence of $k_BT_c/w_T$ with
$w_L/w_T$ obtained from simulations. As in Section II, we restrict
the study to $\theta=1/3 (2/3)$ and $-1<w_L/w_T<1$.

\vspace{0.5cm}
\noindent{\bf Case I: $\theta=1/3$ and $w_L/w_T>0$}
\vspace{0.5cm}

 In general, $e_{\infty}(\theta)$ can be calculated from
 mean-field approximation. Thus,

\begin{equation}
e_{\infty}(\theta) = \frac{1}{2M} \left(6 N \theta w_T + 2 N
\theta w_L   \right) \label{u2}
\end{equation}
\noindent In this case $\theta=N/M=1/3$, and

\begin{equation}
e_{\infty}(1/3) = \frac{1}{3} w_T + \frac{1}{9} w_L \label{u1/3}
\end{equation}

In order to calculate the entropy of the disordered state, the
configurational factor of monomers $\Omega$, is employed

\begin{equation}
\Omega= \frac{M!}{N!(M-N)!}   \label{k1}
\end{equation}

Thus,

\begin{equation}
s_{\infty} = \lim_{M \rightarrow \infty} \frac{k_B \ln \Omega}{M}   \label{s2}
\end{equation}

In the particular case of $\theta=1/3$, the entropy per site of the disordered state
results,

\begin{equation}
s_{\infty}(1/3) = -k_B \left( \ln \frac{1}{3} + \frac{2}{3} \ln 2 \right)    \label{s1/3}
\end{equation}

In addition, the mean energy per site and the entropy per site for
the ordered state at $\theta=1/3$ and $T=0$ are $e_0(1/3) =
s_0(1/3) = 0$. Then, the critical temperature depends on the mean
energy and the entropy of the disordered state. From eqs.
(\ref{u1/3}), (\ref{s1/3}) and (\ref{fetc1}), we obtain
$T_c(1/3)$:

\begin{eqnarray}
T_c(1/3)  & \approx & \frac{e_{\infty}(1/3)}{s_{\infty}(1/3) } \nonumber \\
& \approx & \frac{\frac{1}{3} w_T + \frac{1}{9} w_L}{-k_B \left( \ln \frac{1}{3} + \frac{2}{3} \ln 2 \right) }
\label{tc1/3}
\end{eqnarray}
Finally,
\begin{equation}
\frac{k_B T_c(1/3)}{w_T}  \approx  \frac{\frac{1}{3} + \frac{1}{9} \frac{w_L}{w_T}}{-\ln \frac{1}{3}
- \frac{2}{3} \ln 2 }
\ \ \ \ \ \ \ \ \left(w_T > 0 \ \ {\rm and} \ \ w_L/w_T>0 \right)  \\
\label{tc1/3a}
\end{equation}

\vspace{0.5cm}
\noindent{\bf Case II: $\theta=1/3$ and $w_L/w_T<0$}
\vspace{0.5cm}

 The mean energy and the entropy of the disordered
system are as in eqs. (\ref{u2}) and (\ref{s2}). On the other
hand, for the ordered system, $s_0(1/3) = 0$ and $e_0(1/3)=w_L/3$.
Then,

\begin{equation}
\frac{k_B T_c(1/3)}{w_T}  \approx  \frac{\frac{1}{3} - \frac{2}{9} \frac{w_L}{w_T}}{-\ln \frac{1}{3}
- \frac{2}{3} \ln 2 }
\ \ \ \ \ \ \ \ \left(w_T > 0 \ \ {\rm and} \ \ w_L/w_T<0 \right)  \\
\label{tc1/3b}
\end{equation}

As it is expected, the calculations for $\theta=2/3$ (do not shown here)
provide identical results as Cases I and II.

Fig. 6 shows the comparison between the simulated results
previously presented in Table II and the theoretical predictions
obtained from FEMCA for the critical temperature as a function of
$w_L/w_T$. The MC simulations reveal the main characteristics for
the behavior of the critical temperature versus $w_L/w_T$: $i)$
the curve presents a minimum for $w_L/w_T=0$; and $ii)$ for
negative values of $w_L/w_T$, the critical temperatures are higher
than the corresponding ones for positive $w_L/w_T$'s. Both
characteristics are very well reproduce by FEMCA.

The physical meaning of the main features of the critical
temperature can be interpreted from the theoretical approach. In
this framework, the eq. (\ref{fetc1}) shows that $k_B T_c/w_T$
depends on the mean energy and the entropy of the disordered
state. The behavior of these quantities as a function of $w_L/w_T$
allows to understand the arguments presented in the previous
paragraph. Thus, the values of $s_{\infty}$, $s_{0}$ and
$e_{\infty}$ are identical for repulsive and attractive
longitudinal interactions. In addition, the magnitude of $e_{0}$
is constant ($e_0=0$) for $w_L>0$ and increases with $w_L$
($e_0=w_L/3$) for $w_L<0$. Now we can interpret the difference
between the two regimes in Fig. 6. From $-1<w_L/w_T<1$, the
variation in the entropy is constant. On the other hand, the
variation of the mean energy increases linearly with $w_L$, being
higher for $w_L<0$ than for $w_L>0$.

\section{Conclusions}\label{conclusions}

In the present work, we have addressed the critical properties of
a simple lattice-gas model, which mimics a nanoporous environment,
where each nanotube or unit cell is represented by a
one-dimensional array. The results were obtained by using MC
simulations, finite-size scaling theory and the recently reported
FEMCA, which is based on a free energy minimization
criterion.

The system was characterized by two parameters $w_L/w_T$ and $k_BT/w_T$,
being $w_L$ and $w_T$, the longitudinal and transversal energy, respectively.
We focused on the case of repulsive transversal interaction energy
among adsorbed particles ($w_T = 1$) and $\theta=1/3 (2/3)$, in such a way
that a rich variety of ordered phases are observed in the adlayer:

\begin{itemize}

\item For $w_L/w_T=0$, the system is equivalent to the well-known
triangular lattice in 2D.

\item For $w_L/w_T<0$, the formation of pairs of nearest-neighbor
adsorbed particles along the nanotubes is favored. Consequently,
the $(\sqrt{3} \times \sqrt{3})$ and $(\sqrt{3} \times
\sqrt{3})^*$ phases are reinforced and extend along the channels.
The critical temperature decreases from $0.7817(1)$ for
$w_L/w_T=-1$ to $0.3354(1)$ for $w_L/w_T=0$.

\item For $w_L/w_T>0$, the $(\sqrt{3} \times \sqrt{3})$ and
$(\sqrt{3} \times \sqrt{3})^*$ structures are formed in the planes
at low-temperatures and order along the channels in a array of
alternating particles. The critical temperature increases from
$0.3354(1)$ for $w_L/w_T=0$ to $0.6098(5)$ for $w_L/w_T=1$.

\end{itemize}

With respect to the analytical approach, FEMCA provides results in very good qualitative
agreement with MC simulations and constitutes a theoretical framework in order to interpret
the behavior of $k_BT_c/w_T$ vs $w_L/w_T$ in the critical concentrations.

Future efforts will be directed to (a) include attractive $w_T$
longitudinal interactions between the adparticles, (b) obtain the
phase diagram $k_BT_c/w_T$ versus $\theta$ in the whole range of
coverage, and (c) develop an exhaustive study on critical
exponents and universality.

\subsection*{Acknowledgment}

This work was supported in part by CONICET (Argentina) under
project PIP 6294 and the Universidad Nacional de San Luis
(Argentina) under the projects 328501 and 322000. One of the
authors (AJRP) is grateful to the Departamento de Qu\'{\i}mica,
Universidad Aut\'onoma Metropolitana-Iztapalapa (M\'exico, D.F.)
for its hospitality during the time this manuscript was prepared.

\newpage

\section{Table and Figure Captions}
\begin{itemize}

\item[Table I] Parameters of the simulated tempering runs for two
typical cases ($w_L/w_T=-1, 1$).

\item[Table II] Critical temperatures corresponding to the
critical coverage $\theta=1/3(2/3)$. The data were obtained from
the crossing of the cumulants.

\item[Figure 1:] a) Snapshot of two successive planes, $k$ and
$k+1$, for a possible configuration of the ordered phase appearing
at $\theta=1/3$ and repulsive longitudinal interactions. Solid
circles represent occupied sites.  The different sublattices used
in order to define a local order parameter characterizing this low
temperature structure are shown in parts b)-c).

\item[Figure 2:] Different sublattices defined for attractive
longitudinal interactions and $\theta=1/3$.

\item[Figure 3:] $U_R(T)$ and $U_L(T)$ [inset] versus $k_BT/w_T$,
for a typical case of $w_L < 0$: $\theta =1/3$ and $w_L/w_T=-1$.

\item[Figure 4:] $U_R(T)$ versus $k_BT/w_T$ for $\theta =1/3$ and
$w_L/w_T=1$.

\item[Figure 5:] $U_L(T)$ versus $k_BT/w_T$ for $\theta =1/3$ and
$w_L/w_T=-1$.

\item[Figure 6:] Comparison between simulated and theoretical
results for $k_BT_c/w_T$ vs. $w_L/w_T$ at $\theta=1/3(2/3)$. The
dotted lines are a guide for the eyes.
\end{itemize}

\newpage
\begin{center}
\vspace{1.5cm} \noindent {\bf {TABLE I}} \vspace{0.5cm}
$$
\begin{array}{|c|c|c|c|c|c|c|c|c|}
\hline  \hline
w_L/w_T & R & L & m &  n_1 &  n_2 & n_{MCS} & k_BT_{min}/w_T & k_BT_{max}/w_T \\
\hline
   & 18 & 60 & 12 & 10^3  & 10^5  & 10^5 & 0.780053 & 0.782705 \\
\cline{2-9}
-1  & 24 & 60 & 20 & 10^3  &   10^5  & 10^5  & 0.777400 & 0.784600 \\
\cline{2-9}
  &  36 & 60 & 12 &  10^3 &   10^5  & 10^5  &  0.780053 & 0.782705  \\
\cline{2-9}
  &  48 & 60 & 11 & 10^3  &   10^5  & 10^5  & 0.780500 & 0.782500 \\
\hline \hline
   & 18 & 60 & 12 &  10^3 &   5.10^5  & 5.10^5  & 0.601053 & 0.617632 \\
\cline{2-9}
 1 & 24 & 60 & 12 & 10^3  &   5.10^5  & 5.10^5  &  0.601053 & 0.617632  \\
\cline{2-9}
  &  36 & 60 & 12 & 10^3  &   5.10^5  & 5.10^5  &  0.601053 & 0.617632  \\
\cline{2-9}
  &  48 & 60 & 12 & 10^3  &   5.10^5  & 5.10^5  &  0.601053 & 0.617632  \\
\hline \hline
\end{array}
$$

\vspace{1.5cm}
\noindent {\bf {TABLE II}}
\vspace{0.5cm}

\begin{tabular}{|c|c|}            \hline  \hline

 & \multicolumn{1}{c|}{$k_BT_c/w_T$} \\ \cline{2-2}

$w_L/w_T$ & $\theta=1/3~(2/3)$  \\  \hline

-1.00   & 0.7817(1)       \\ \hline

-0.75  & 0.695(4)    \\  \hline

-0.50  & 0.604(4)    \\  \hline

-0.25  & 0.495(4)    \\  \hline

0.00  & 0.3354(1)    \\  \hline

0.25  & 0.436(1)    \\  \hline

0.50  & 0.506(1)    \\  \hline

0.75  & 0.562(1)    \\  \hline

1.00  & 0.6098(5)   \\  \hline

\hline
\end{tabular}
\end{center}
\end{large}


\begin{center}

\begin{figure}
\includegraphics[width=10cm,clip=true]{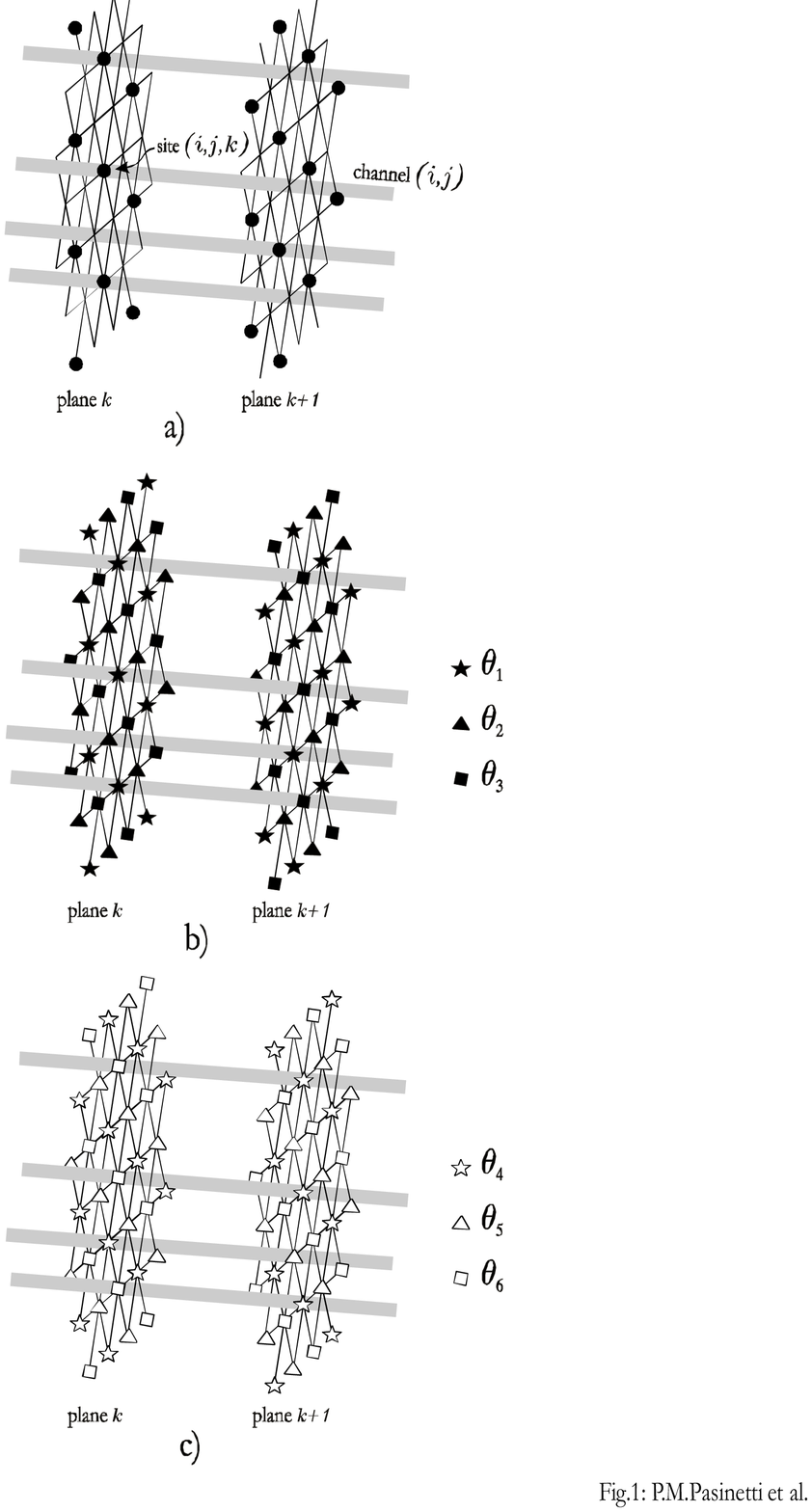}
\end{figure}

\begin{figure}
\includegraphics[angle=-90,width=10cm,clip=true]{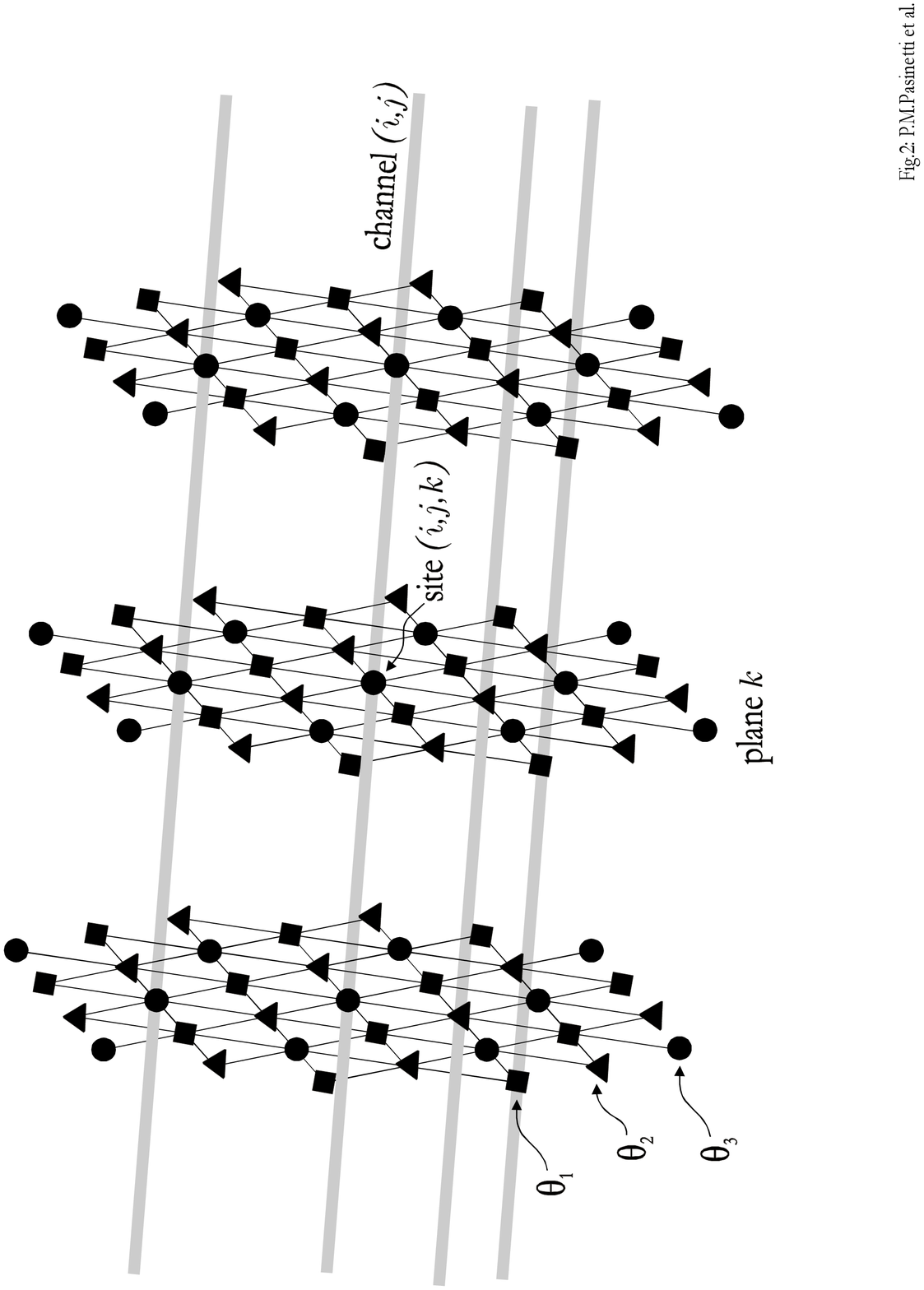}
\end{figure}

\begin{figure}
\includegraphics[angle=90,width=18cm,clip=true]{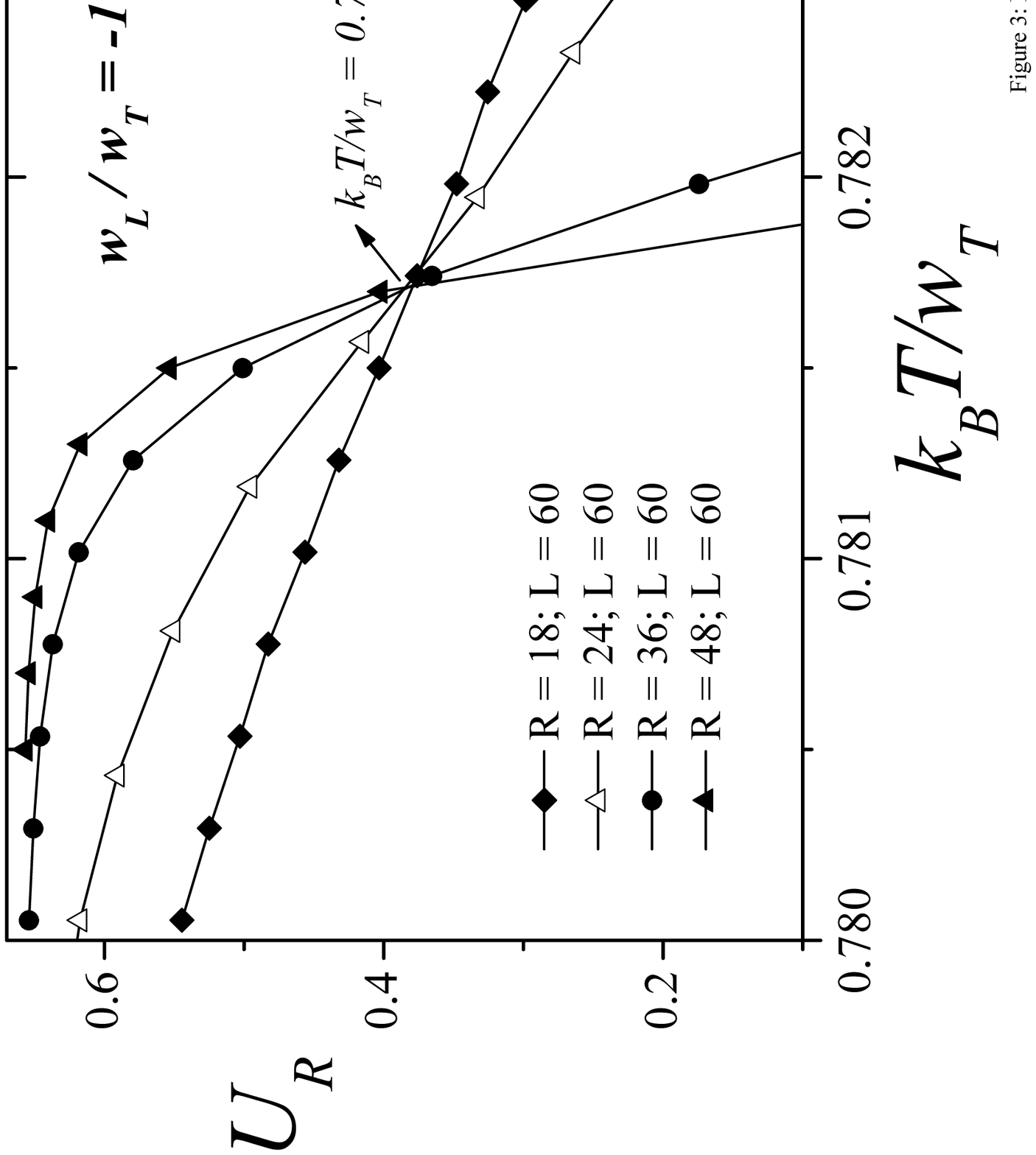}
\end{figure}

\begin{figure}
\includegraphics[angle=90,width=18cm,clip=true]{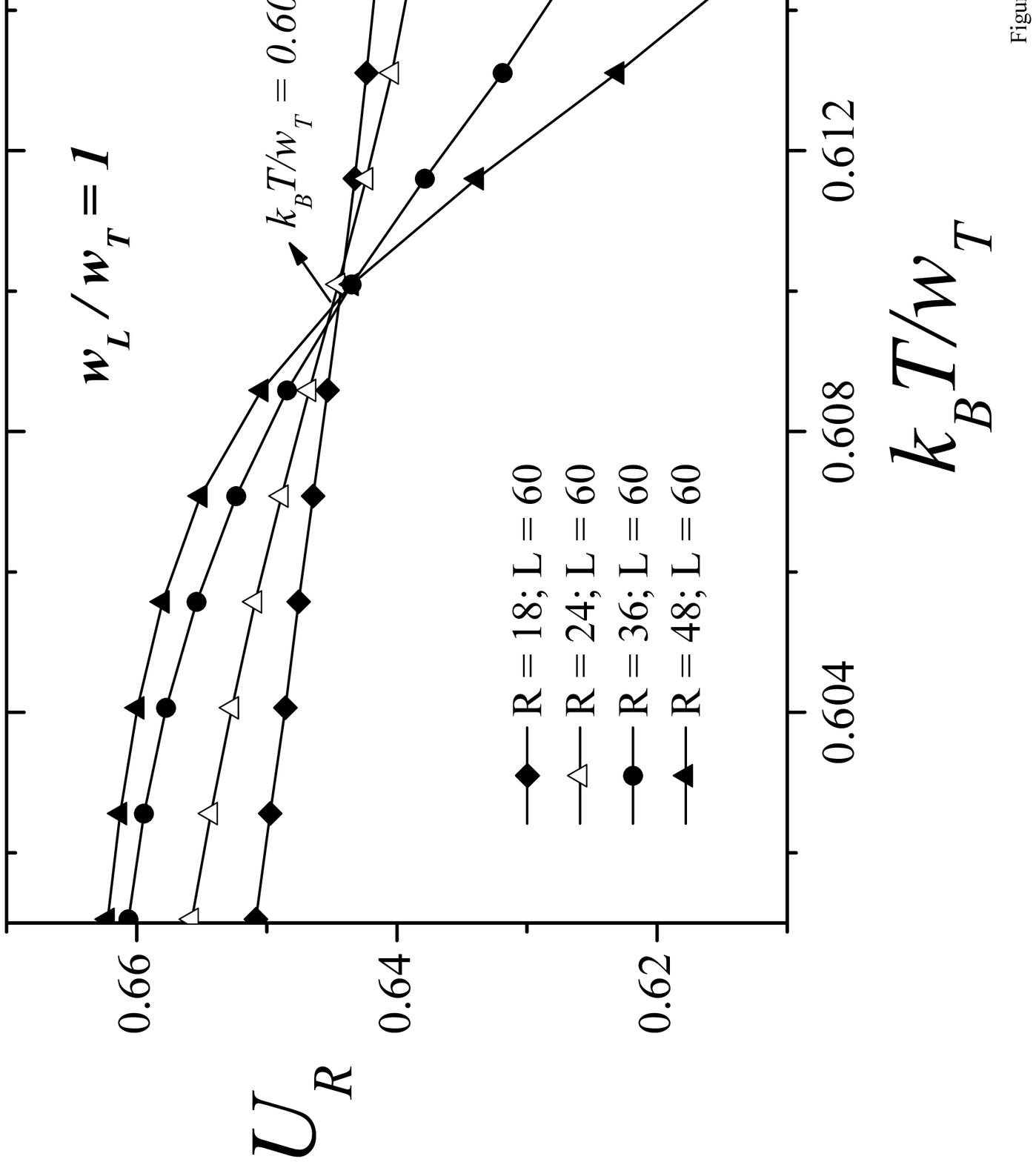}
\end{figure}

\begin{figure}
\includegraphics[angle=90,width=18cm,clip=true]{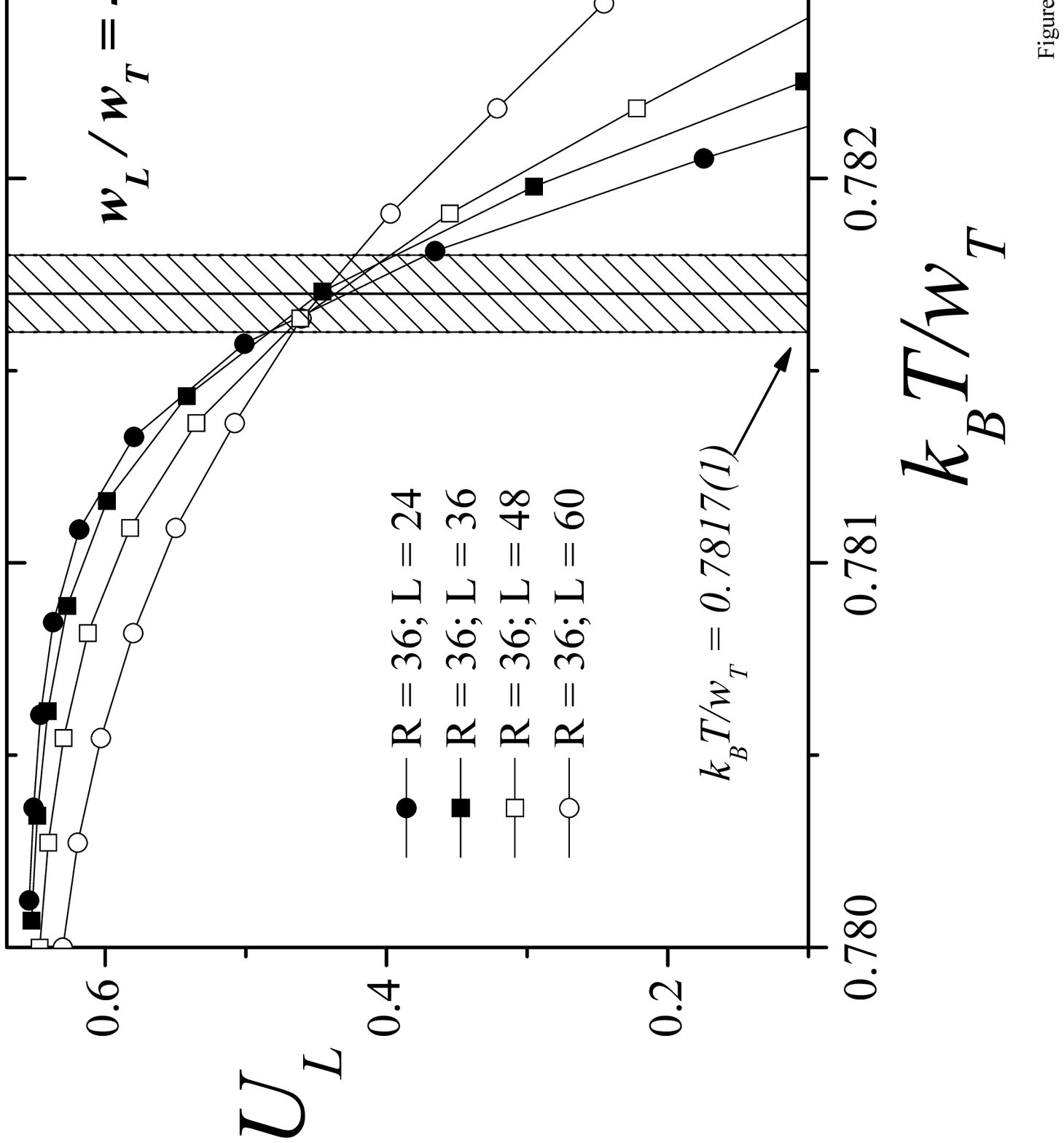}
\end{figure}

\begin{figure}
\includegraphics[angle=90,width=18cm,clip=true]{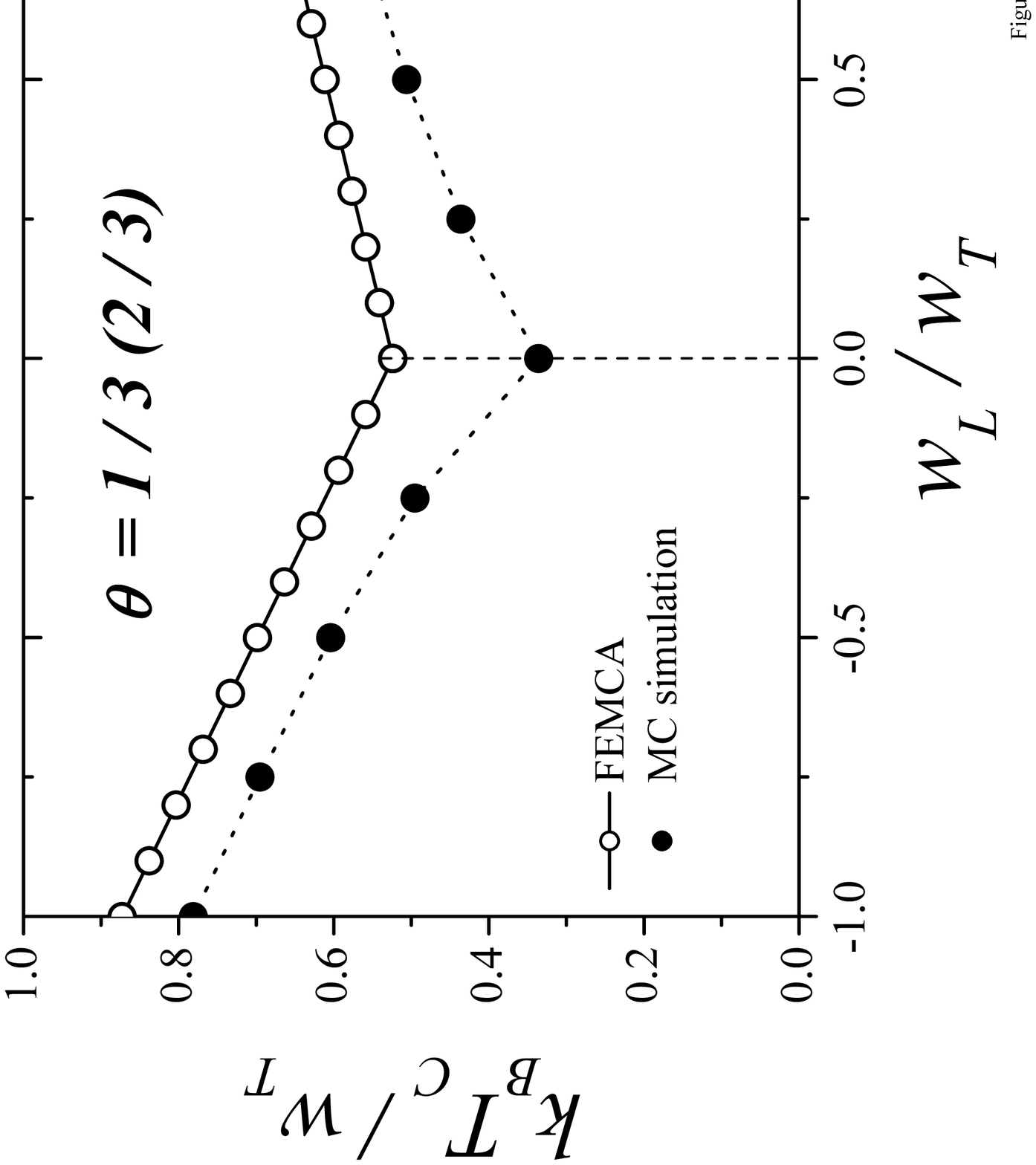}
\end{figure}

\end{center}



\newpage

\begin{thebibliography}{10}
\bibitem{Fowler} R. H. Fowler and E. A. Guggenheim, {\it Statistical
Thermodynamics} (Cambridge University Press, Cambridge, 1949).
\bibitem{Hill} T. L. Hill, {\it An Introduction to Statistical Thermodynamics} (Addison Wesley Publishing Company,
Reading, MA, 1960).
\bibitem{Stanley} H. E. Stanley, {\it Introduction to Phase Transitions and Critical Phenomena}
 (Oxford University Press, New York, 1971).
\bibitem{Langmuir} I. Langmuir, J. Am. Chem. Soc. {\bf 34}, 1310 (1912); {\bf 37}, 417 (1951);
{\bf 54}, 2798 (1932); Gen. Electron. Rev. {\bf 29}, 153 (1926);
I. Langmuir and K. H. Kingdom, Proc. R. Soc. Lond. A {\bf 107}, 61
(1925); I. Langmuir and K. H. Kingdom, Phys. Rev. {\bf 34}, 129
(1919); I. Langmuir and J. B. Taylor, Phys. Rev. {\bf 44}, 423
(1933); I. Langmuir and D. S. Villars, J. Am. Chem. Soc. {\bf 53},
486 (1931).
\bibitem{Ising} E. Ising, Z. Phys. {\bf 31}, 253 (1925).
\bibitem{Kramers} H. A. Kramers and G. H. Wannier, Phys. Rev. {\bf 60}, 252 (1941);
Phys. Rev. {\bf 60}, 263 (1941).
\bibitem{Montroll} E. Montroll, J. Chem. Phys. {\bf 9}, 706 (1941).
\bibitem{Onsager} L. Onsager, Phys. Rev. {\bf 65}, 117 (1944).
\bibitem{Binder} K. Binder and D. P. Landau, Phys. Rev. B {\bf 21}, 1941 (1980).
\bibitem{Landau} D. P. Landau, Phys. Rev. B {\bf 27},  5604 (1983).
\bibitem{Phares} A. J. Phares and F. J. Wunderlich, J. Math. Phys. {\bf 26}, 2491 (1985); Phys. Rev. E {\bf 52}, 2236 (1995);
 Phys. Rev. E {\bf 55}, 2403 (1997);  Surf. Sci. {\bf 425}, 112 (1999);  Surf. Sci. {\bf 452}, 108 (2000);
Surf. Sci. {\bf 479}, 43 (2001).
\bibitem{PRE2} F. Rom\'a and A. J. Ramirez-Pastor, Phys. Rev. E {\bf 69}, 036124 (2004).
\bibitem{Patry} A. Patrykiejew, S. Sokolowski and K. Binder, Surf. Sci. Rep. {\bf 37}, 207 (2000).
\bibitem{JCP1PRE1} F. Bulnes, A. J. Ramirez-Pastor and G. Zgrablich, J. Chem. Phys. {\bf 115}, 1513 (2001);
Phys. Rev. E {\bf 65}, 31603 (2002).
\bibitem{Nitta} T. Nitta, M. Kuro-Oka and K. Katayama, J. Chem Eng. Japan {\bf 17}, 45 (1984).
\bibitem{Rudzi} W. Rudzinki, K. Nieszporek, J.M. Cases, L.I. Michot and
              F. Villeras, Langmuir {\bf 12}, 170 (1996).
\bibitem{Boro1} M. Bor\'owko and W. R\.zysko, Journal of Colloid
and Interface Science \bf 182, \rm 268 (1996); Ber. Bunsenges.
Phys. Chem. \bf 101, \rm 84 (1997).
\bibitem{Boro2} W. R\.zysko and M. Bor\'owko, Journal of Chemical Physics
 \bf 117, \rm 4526 (2002); Surface Science \bf 520, \rm 151 (2002).
\bibitem{PRB3} A.J. Ramirez-Pastor, T.P. Eggarter, V.D. Pereyra, and J.L. Riccardo,
Phys. Rev. B {\bf 59}, 11027 (1999).
\bibitem{SS3LANGMUIR5} A. J. Ramirez-Pastor,  J. L. Riccardo and V. D. Pereyra, Surf. Sci. {\bf 411}, 294 (1998);
Langmuir {\bf 16}, 10169 (2000).
\bibitem{LANGMUIR6} F. Rom\'a, A. J. Ramirez-Pastor and J. L. Riccardo,
Langmuir {\bf 16}, 9406 (2000); J. of Chem. Phys. {\bf 114}, 10932
(2001).
\bibitem{PRB4} F. Rom\'a, A. J. Ramirez-Pastor and J. L. Riccardo,
Phys. Rev. B {\bf 68}, 205407 (2003).
\bibitem{Ijima} S. Ijima, Nature {\bf 354}, 56 (1991).
\bibitem{Ijima1} S. Ijima and T. Ichihashi, Nature {\bf 363}, 603 (1993).
\bibitem{Bethune} D.S. Bethune, C.H. Kiamg, M. S. deVries, G. Gorman, R. Savoy,
J. Vasquez and R. Beyers, Nature {\bf 363}, 605 (1993).
\bibitem{Ajayan} P.M. Ajayan and S. Ijima, Nature {\bf 361}, 333 (1993).
\bibitem{Ebbensen} E. Dujardin, T. W. Ebbesen, H. Hiura and K. Tanigaki, Science {\bf 265}, 1850 (1994).
\bibitem{Martin} C. Martin, J.P. Coulomb, Y. Grillet and R. Kahn,
{\it Fundamentals of Adsorption: Proceedings of the Fifth
International Conference}, edited by M.D. LeVan (Kluwer Academic
Publishers, Boston Massachusetts, 1996), p.587.
\bibitem{Gubbins} R. Radhakrishnan and K.E. Gubbins, Phys. Rev. Lett. {\bf 79}, 2847 (1997).
\bibitem{Boutin} A. Boutin, R.J.-M. Pellenq and D. Nicholson, Chem. Phys. Lett. {\bf 219}, 484 (1994).
\bibitem{Lachet} V. Lachet, A. Boutin, R.J.-M. Pellenq, D. Nicholson and A.H. Fuchs,
J. Phys. Chem. {\bf 100}, 9006 (1996).
\bibitem{Maris} T. Maris, T.J.H. Vlugt and B. Smit, J. Phys. Chem. {\bf 102}, 7183 (1998).
\bibitem{Martin1} C. Martin, N. Tosi-Pellenq, J. Patarin and J.P. Coulomb, Langmuir {\bf 14}, 1774 (1998).
\bibitem{Urban} N. M. Urban, S. M. Gatica, M.W. Cole and J.L. Riccardo, Phys. Rev. B {\bf 71}, 245410 (2005).
\bibitem{Calbi} M. M. Calbi and J.L. Riccardo, Phys. Rev. Lett. {\bf 94}, 246103 (2005).
\bibitem{Gelb} Lev D. Gelb, K. E. Gubbins, R. Radhakrishnan and M. Sliwinska-Bartkowiak,
Rep. Prog. Phys. {\bf 62}, 1753 (1999).
\bibitem{Cole1} R.A. Trasca, M.M. Calbi and M.W. Cole, Phys. Rev. E {\bf 65}, 061607 (2002).
\bibitem{Cole} R.A. Trasca, M.M. Calbi, M.W. Cole and J.L. Riccardo, Phys. Rev. E {\bf 69}, 011605 (2004).
\bibitem{entropy} P. M. Pasinetti, J. L. Riccardo and A. J. Ramirez-Pastor,
J. Chem. Phys. {\bf 122}, 154708 (2005).
\bibitem{isotherm} P. M. Pasinetti, J. L. Riccardo and A. J. Ramirez-Pastor,
 Physica A {\bf 355}, 383 (2005).
\bibitem{Hukushima} K. Hukushima and K. Nemoto, J. Phys. Soc. Jpn. {\bf 65}, 1604 (1996).
\bibitem{Earl} D. J. Earl and M. W. Deem, e-print cond-mat/0508111.
\bibitem{Fisher} M. E. Fisher, in: {\it Critical Phenomena}, edited by M. S. Green,
(Academic Press, London, 1971) pp.1.
\bibitem{Privman} V. Privman, {\it Finite Size Scaling and Numerical Simulation of
             Statistical Systems} (World Scientific, Singapore, 1990).
\bibitem{Binder2} K. Binder, {\it Applications of the Monte Carlo Method in Statistical
Physics: Topics in current Physics} (Springer, Berlin, 1984), Vol.
36.
\bibitem{Nicholson} D. Nicholson and N. G. Parsonage, {\it Computer Simulation
and the Statistical Mechanics of Adsorption} (Academic Press,
London, 1982).
\bibitem{Metropolis}  N. Metropolis, A.W. Rosenbluth, M.N. Rosenbluth, A.H. Teller
and  E. Teller, J. Chem. Phys. {\bf 21}, 1087 (1953).
\bibitem{Kawasaki} K. Kawasaki, in: {\it Phase Transitions and Critical Phenomena},
edited by C. Domb and M. S. Green, (Academic Press, London, 1972),
Vol. 2.
\bibitem{foot1} By inspecting the Fig. 1, it's possible to
note that the configuration presented in a) was formed on the
sublattice $1$.
\bibitem{Metcalf}  B. D. Metcalf, Phys. Lett. A {\bf 45}, 1 (1973).
\bibitem{Kinzel} W. Kinzel and M. Schick, Phys. Rev. B {\bf 23}, 3435 (1981).
\bibitem{Chin} K. K. Chin and D. P. Landau, Phys. Rev. B {\bf 36}, 275 (1987).
\bibitem{Triang} P. M. Pasinetti, F. Rom\'a, J. L. Riccardo and A. J. Ramirez-Pastor, e-print cond-mat/0606391.
\bibitem{foot2} As it is common in MC simulations, one represents the
system with a unit cell of sites that is repeated periodically.
Then, the choice of appropriate sizes in the transversal direction
has to be done in such a way that the ordered structures are not
disturbed.
\bibitem{Barouch} E. Barouch, B. M. McCoy and T. T. Wu, Phys. Rev. Lett. {\bf 31}, 1409 (1973).
\bibitem{BinderWang} K. Binder and J. S. Wang, J. Stat. Phys. {\bf 55}, 87 (1989).
\bibitem{foot3} A systematic analysis of critical exponents for each
$w_L/w_T$ was not carried out since this was out of the scope of
the present work.
\bibitem{foot4} The calculations for
$w_L/w_T=1.0$ and $w_L/w_T=-1.0$ were carried out with an effort
reaching almost the limits of our computational capabilities. In
the case of intermediate values of $w_L/w_T$, the number of MCS
was restricted in order to get results in a reasonably
computational time. These conditions are reflected in the
different values of the numerical errors reported in Table II.
\end{thebibliography}
\end{document}